\documentclass{article}

     \PassOptionsToPackage{numbers, compress}{natbib}

     \usepackage[preprint]{neurips_2024}

\usepackage[utf8]{inputenc} 
\usepackage[T1]{fontenc}    
\usepackage{hyperref}       
\usepackage[final]{changes} 
\usepackage{url}            
\usepackage{wrapfig}
\usepackage{booktabs}       
\usepackage{amsfonts,bm}       
\usepackage{nicefrac}       
\usepackage{microtype}      
\usepackage{xcolor}         
\usepackage{ulem}
\usepackage{picins}
\usepackage[inline, shortlabels]{enumitem}
\usepackage{todonotes}
\usepackage{changes}
\usepackage{float,subcaption}
\usepackage{caption}
\usepackage{todonotes}
\usepackage{makecell}
\usepackage{tabulary}
\usepackage{tabularx}
\usepackage{graphicx,scalerel}
\usepackage{stackengine}
\usepackage{adjustbox}
\usepackage{diagbox}
\usepackage{tabstackengine}
\usepackage{amsmath,bm}
\usepackage{multirow}
\usepackage{footnote}

\newcommand\sbullet[1][.5]{\mathbin{\ThisStyle{\vcenter{\hbox{%
  \scalebox{#1}{$\SavedStyle\bullet$}}}}}%
}

\definecolor{ao(english)}{rgb}{0.0, 0.5, 0.0}

\definechangesauthor[color=purple, name={jean-patrick}]{JP}
\definechangesauthor[color=violet, name={Milad}]{ML}

\title{NeurIPS 2024 ML4CFD Competition: Harnessing Machine Learning for Computational Fluid Dynamics in Airfoil Design}

\author{
     Mouadh Yagoubi, David Danan,  Milad Leyli-Abadi, Jean-Patrick Brunet, Maroua Gmati \\
     IRT SystemX, Palaiseau, France \And  Ahmed Mazari $^1$, Florent Bonnet $^{1,2}$ \\ \small{1}. SimAI team, Ansys Inc, France \\ \small{2}. Sorbonne Université, CNRS, ISIR \\ \And Asma Farjallah \\ NVIDIA  \And Paola Cinnella \\ Institut Jean Le Rond D’Alembert\\Sorbonne Université, France \And Patrick Gallinari \\ Sorbonne Université, CNRS, ISIR \\
Criteo AI Lab \And Marc Schoenauer \\ INRIA Saclay, France 
} 
\newcommand*{\myfullcircle}[1]{
    \begin{tikzpicture}[scale=0.1, line width=.3mm]]%
    \draw[black, fill=#1] (0,0) circle (1.5);
    \end{tikzpicture}
}

\begin{document}

\maketitle



\begin{abstract}

The integration of machine learning (ML) techniques for addressing intricate physics problems is increasingly recognized as a promising avenue for expediting simulations. However, assessing ML-derived physical models poses a significant challenge for their adoption within industrial contexts. This competition is designed to promote the development of innovative ML approaches for tackling physical challenges, leveraging our recently introduced unified evaluation framework known as Learning Industrial Physical Simulations (LIPS). Building upon the preliminary edition held from November 2023 to March 2024\footnote{\href{https://www.codabench.org/competitions/1534/}{https://www.codabench.org/competitions/1534/}}, this iteration centers on a task fundamental to a well-established physical application: airfoil design simulation, utilizing our proposed AirfRANS dataset. The competition evaluates solutions based on various criteria encompassing ML accuracy, computational efficiency, Out-Of-Distribution performance, and adherence to physical principles. Notably, this competition represents a pioneering effort in exploring ML-driven surrogate methods aimed at optimizing the trade-off between computational efficiency and accuracy in physical simulations. Hosted on the Codabench platform, the competition offers online training and evaluation for all participating solutions.
\end{abstract}

\subsection*{Keywords}
 · Geometric Deep Learning · Hybridization · Benchmark · PDE  · CFD 

\section{Competition description}


\subsection{Background and impact}

Nowadays, numerical simulation has become indispensable for designing and managing intricate physical systems due to its cost-effectiveness compared to real-world experiments. While classical numerical methods often offer accurate predictions of system behavior, they typically come with a high computational cost, limiting their applicability in complex industrial settings. Machine learning (ML) approaches, proven successful in diverse domains like computer vision, natural language processing, and speech recognition, are increasingly gaining traction. Recently, they gain a particular attention in physical domains where traditional numerical methods face challenges or involve expensive and imprecise computations.

Deep Learning (DL) techniques, in particular, have garnered attention for their ability to tackle complex tasks, showing promising results across various physical domains (see e.g.,\cite{tompson2016,kasim2021building,rasp_deep_2018,sanchez2020learning, menier2023cd}). These approaches facilitate significant acceleration of simulations by replacing certain computational components with data-driven models. The primary goal of the presented challenge is to foster the development of novel ML solutions for solving physical problems. To achieve this, we propose focusing on a well-established Computational Fluid Dynamics (CFD) use case: airfoil design. Our aim is to establish an efficient benchmarking framework for evaluating submitted solutions.

In addition to traditional machine learning metrics assessing accuracy and speedup, we aim to incorporate other essential criteria crucial for validating ML-based physical models in industrial settings. These include Out-of-Distribution (OOD) generalization and physics compliance. To facilitate this evaluation, we introduce our recently developed benchmarking platform, LIPS (Learning Industrial Physical Simulation)\cite{leyli2022lips}, which is briefly outlined in the following paragraph.

\begin{figure}[H]
    \centering
    \includegraphics[width=0.8\linewidth]{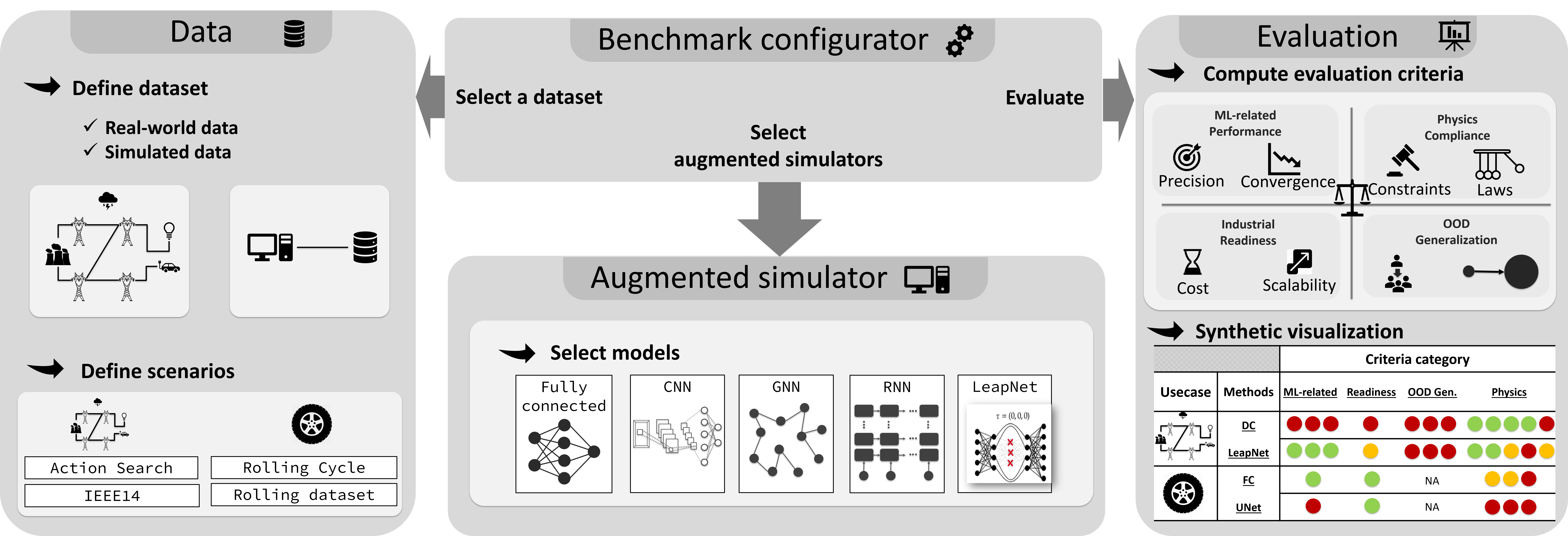}
    \caption{Learning Industrial Physical Simulation (LIPS) framework}
    \label{fig:lips}
\end{figure}

\paragraph{LIPS Framework} \label{sec: LIPS}
The LIPS Framework \cite{leyli2022lips}  serves as a unified and extensible platform designed for benchmarking ML-based physical simulations in a uniform yet adaptable manner. It facilitates the evaluation of various ML-based physical simulators through four key modules: data management, benchmark configurator, augmented simulator, and evaluation (Figure \ref{fig:lips}).

Primarily, the LIPS framework is utilized to establish generic and comprehensive evaluation criteria for ranking submitted solutions. This evaluation process encompasses multiple aspects to address industrial requirements and expectations. The following categories outline the evaluation criteria considered:
\begin{itemize}
    \item  \textbf{ML-related performance}: we focus on assessing the trade-offs between typical model accuracy metrics like Mean Squared Error (MSE) and computation time.
    \item \textbf{Out-of-Distribution (OOD) generalization}: given the necessity in industrial physical simulation to extrapolate over minor variations in problem geometry or physical parameters, we incorporate OOD geometry evaluation, such as unseen airfoil mesh variations.
    \item \textbf{Physics compliance}: ensuring adherence to physical laws is crucial when simulation results influence real-world decisions. Depending on the benchmark's criticality, this category of criteria aims to specify the types and quantity of physical laws that must be satisfied.
\end{itemize}

Note that the Airfoil design task being considered for this competition has already been implemented within the LIPS framework and was successfully utilized during the preliminary edition of this competition.

\subsection{Novelty}
\label{sec:novelty}
The integration of machine learning (ML) with physics represents an expanding area of research, with a multitude of industrial applications. A considerable portion of this research capitalizes on the latest advancements in ML to address challenges encountered in the physical sciences, as exemplified by the recent \textit{Machine Learning and the Physical Sciences} workshop \cite{ml4ps2022}. Only a limited selection of competitions focus on specific industrial use cases within the physical domain, such as molecular simulations (Open Catalyst Challenge \cite{open_catalyst}), particle physics \cite{adam2015higgs}, and robotics \cite{cea_challenge}. Conversely, certain scholarly endeavors, including the ICLR 2023 Workshop on Physics for Machine Learning \cite{physics4ml}, along with studies cited in \cite{equer2023multiscale, chughtai2023neural}, attempt to harness the inherent structures of physical systems and the accumulated body of physics knowledge to forge novel ML methodologies, thereby enhancing our comprehension of these disciplines. \\ What sets this challenge apart is its objective to develop ML strategies specifically designed to tackle physical problems as defined by a particular dataset, the airfoil design task using the AirFrans dataset. \\ We held a trial run of this competition in 2023 \cite{yagoubi2024ml4physim}. The competition facilitated the exploration of novel approaches, including Physics-Informed Neural Networks (PINNs), Graph Neural Networks, Transformers, and combinations with classical ML techniques. The winning strategies, particularly the first-place solution, underscored the potential of combining classical ML with advanced mesh graph treatments \cite{casenave2024mmgp}. \\ Launching a new edition in the scope of NeurIPS will enables the refinement of evaluation criteria to better assess the practical applicability and industrial relevance of solutions, encourages the development of more sophisticated ML models that can address the limitations identified in the previous contest, and strengthens the community by fostering collaboration and knowledge exchange. Moreover, by focusing again on airfoil design but with updated baseline solutions from the preliminary edition (see table \ref{tab:Benchmark_evaluation_circle}), the competition aims to push the boundaries of what's achievable, promoting innovations that could lead to significant advancements in the efficiency and accuracy of physical simulations. To do so, we refine the score computation to make it more challenging, including the consideration of new physical criteria and the assessment of robustness to additive noise.
Besides, submissions will be trained and evaluated on GPU servers provided by NVIDIA as it was the case for  the preliminary competition hosted on Codabench. Having dedicated GPU resources for the challenge allows us to establish a unified process for all participants, ensuring fair evaluation and enabling those without GPU resources to participate.

\subsection{Data}
As mentioned above, the AirfRANS dataset generated for the purpose of the competition is based on specific physical solver that represents the ground truth: OpenFoam (see table \ref{tab:input_output}). For more details about the datasets, the reader could refer to \cite{bonnet2022airfrans}.
For the needs of this challenge, the datasets will be slightly adapted, without any major changes. Each simulation is given as a point cloud defined via the nodes of the simulation mesh, that is to say a discretization of the 2D domain considered.
Inputs and outputs variables for each task are listed in  table \ref{tab:input_output}.
For the challenge, we consider three datasets each comprised of  several samples to be used in  the ML tasks, such as training and testing. Note that each sample in these datasets is associated to a CFD 2D simulation using OpenFoam.
\begin{itemize}
     \item \textbf{Training set}: 103 samples, AirfRANS 'scarce' task, training split, filtered to keep the simulation where the number of reynolds is between 3e6 and 5e6
    \item \textbf{Test set} : 200 samples, AirfRANS 'full' task, testing split
    \item \textbf{OOD test Set} : 496 samples, AirfRANS reynolds task, testing split
\end{itemize}

The decision to employ a limited training split in our approach is driven by a crucial consideration: the high cost associated with acquiring real-world data for tackling industrial challenges. Obtaining authentic samples demands significant investments of time and resources. By incorporating this scarcity into the training set, we aim to closely mirror the constraints faced in practical industrial scenarios. This strategy ensures that the models are trained under conditions that faithfully represent the challenges of real-world deployment, ultimately leading to more effective solutions tailored to industrial needs.

\begin{table}[H]
    \centering
   \caption{Reference data for the task and its related input/output variables.}
    \resizebox{\columnwidth}{!}{
    \begin{tabular}{lllll}
         \toprule
         \textbf{Task} & \textbf{Reference physical simulator} & \textbf{Dataset description} & \textbf{Input variables} & \textbf{Output variables}  \\
         \midrule

         \multirow{2}{*}{Airfoil design} & \multirow{2}{*}{OpenFoam \cite{OpenFOAM}} & AirfRANS  \cite{bonnet2022airfrans} & Positions  & Velocity ($\bar{u}_x$, $\bar{u}_y$)  \\
            & & Documentation \cite{airfrans-doc}          
& Inlet velocity & Pressure divided by the specific mass ($\bar{p_s}$)\\
               & & GitHub \cite{airfrans-github} & Distance to the airfoil & Turbulent kinematic viscosity ($\bar{\nu}_t$)\\
               & & & Normals  &  \\
         \bottomrule
    \end{tabular}}
    \label{tab:input_output}
\end{table}

\subsection{Tasks and application scenarios}

This competition will address the challenge of improving baseline solutions for the airfoil design problem by building ML-based surrogate models. The overall aim is to reduce  the physical simulation cost while preserving acceptable precision for the outputs.  Furthermore, we encourage solutions that can be generalized to solve other scenarios of the airfoils usecase through the OOD generalization dataset. 


\begin{figure}[htb]
  \begin{center}
    \begin{subfigure}{.45\textwidth}
      \centering
      \includegraphics[width=\linewidth]{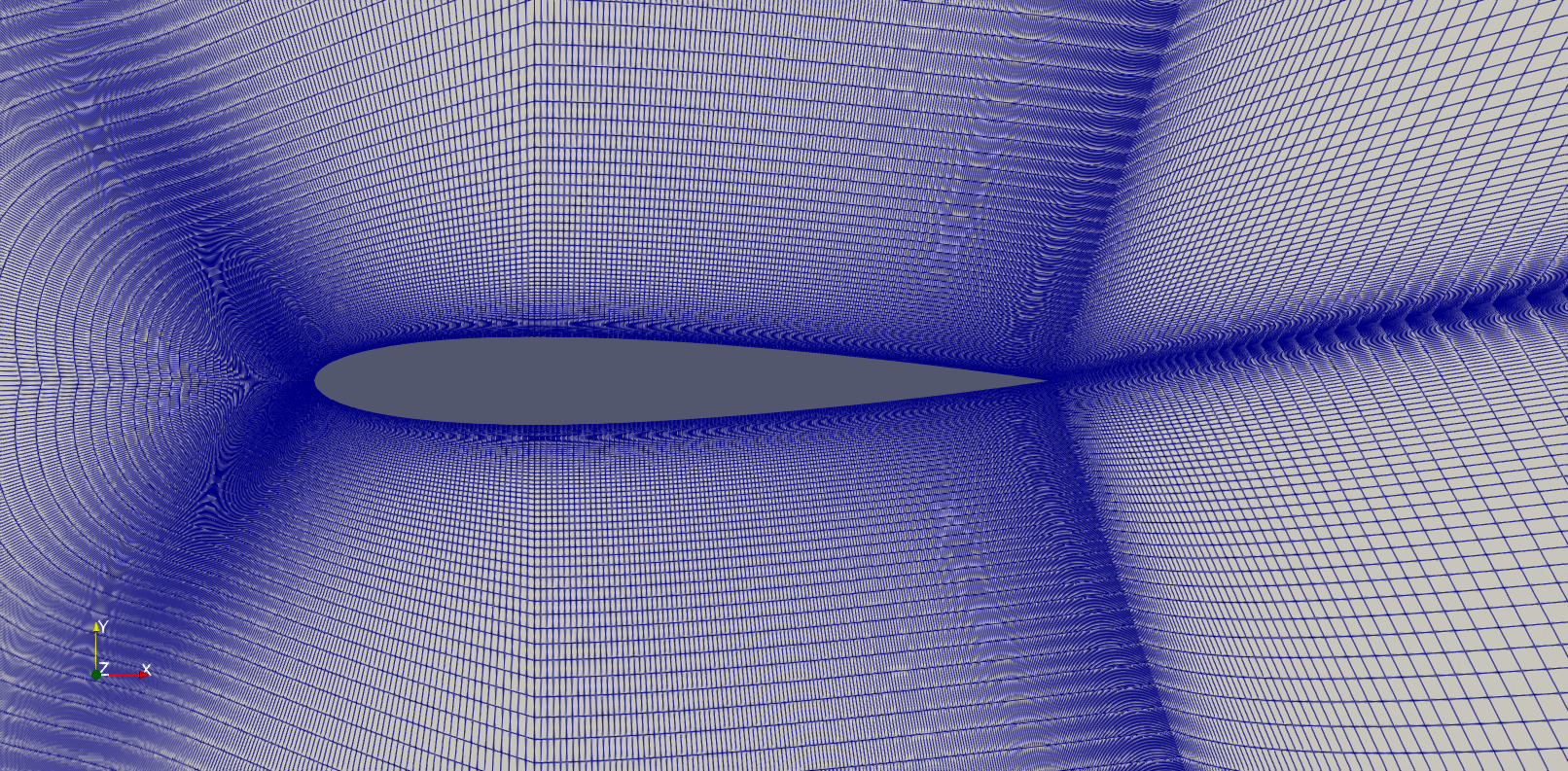} 
      \caption{Input}
    \end{subfigure}
    \begin{subfigure}{.45\textwidth}
      \centering
      \includegraphics[width=\linewidth]{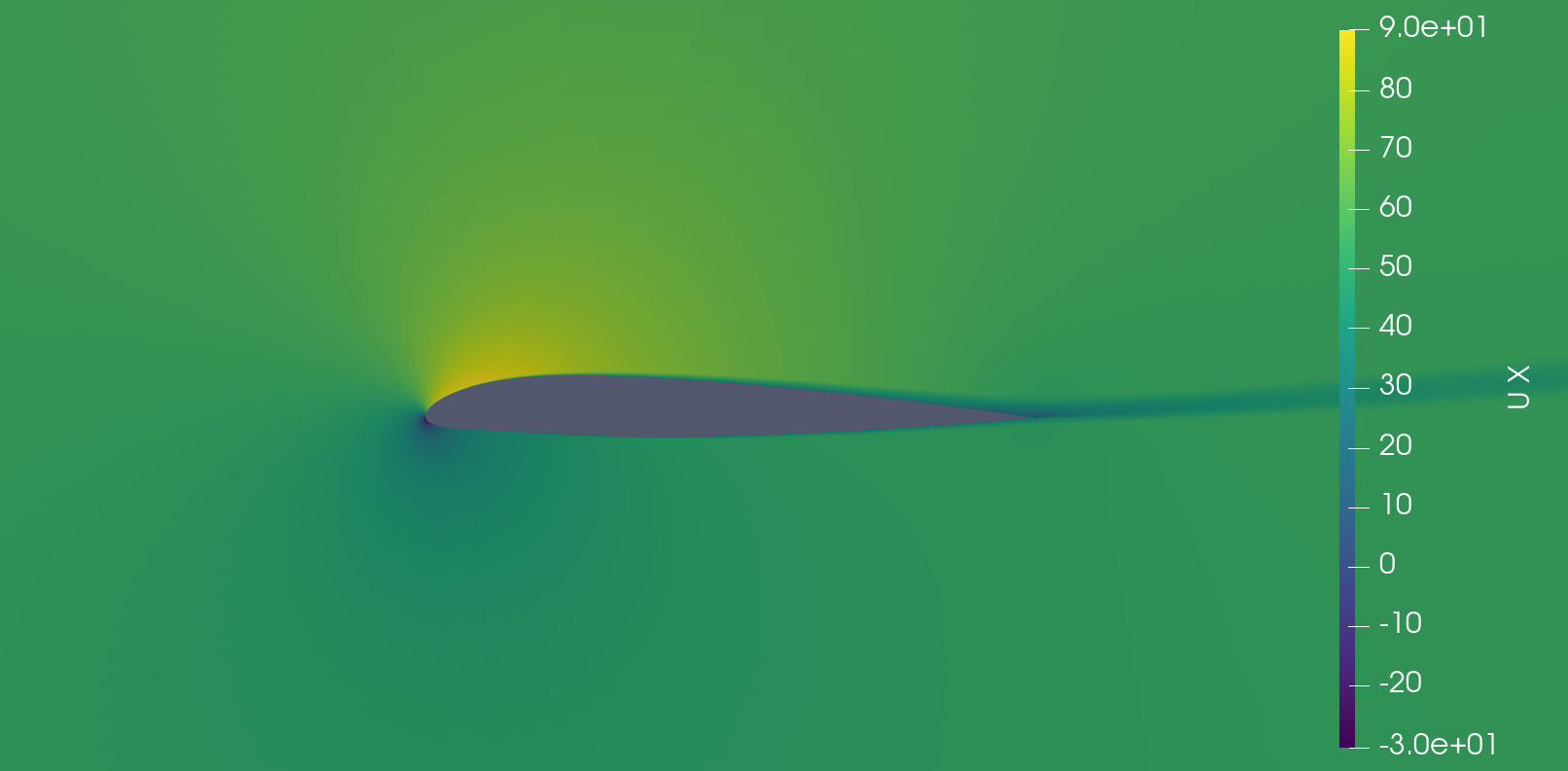}
      \caption{Output}
    \end{subfigure}
    \end{center}

    \caption{(a) Airfoil mesh. (b) $x$-velocity field. 
    }
    \label{fig:cfd_gdl}

\end{figure}

Aircraft design requires a rigorous study of the surrounding aerodynamic fields that could be measured experimentally. However, experiments are extremely costly, time-consuming, and may considerably hinder the time to market new concepts, e.g. to match the stringent constraints on fuel consumption imposed by the energy transition. Moreover, prototypes may not be feasible for highly complex configurations. To circumvent that, one of the solutions is virtual testing, which enables coping with complex constraints and test configurations in a more time-efficient and cheap way.
Hence, numerical simulation is crucial for modeling such physical phenomena that are governed by \textit{Partial Differential Equations} (PDEs) and widely used in \textit{Computational Fluid Dynamics} (CFD) to solve \textit{Navier-Stokes equations} (NS). NS-PDEs are highly nonlinear and their analytical resolution is out of reach.  They are numerically solved with the help of discretization methods such as finite differences, finite elements, or finite volume methods. 
At high Reynolds number (beyond a certain threshold), NS-PDEs involve a complex energy transfer process that cascades from large nearly inviscid length scales to small dissipative ones (\textit{Kolmogorov} microscales), which makes their direct resolution challenging and expensive (thousands of CPU hours). Following this complexity and the related challenges, several strategies are adopted in traditional CFD solvers to solve these equations over different scales, including \textit{Direct Numerical Simulation} (DNS) of all scales, which is prohibitively expensive at high Reynolds, \textit{Large Eddy Simulation} (LES) that models the smaller scales of turbulence and resolves the largest, and \textit{Reynolds-Averaged Navier-Stokes} (RANS) that solves mean field equations while modeling all scales of turbulence.

 Despite the efficiency of these methods, NS-PDEs are still computationally expensive and may take several hours to converge to an accurate solution w.r.t the granularity of meshes. For that reason, we argue that \textit{Deep Learning} (DL) could be a potential surrogate model to cover several CFD tasks including design exploration, design optimization, inverse problem, and real-time control, as well as super-resolution. DL enjoys several advantages: (i) can represent a large family of functions, (ii) can lead to mesh-free models, (iii) trades-off accuracy and complexity, and (iv) enables fast inference once the model is trained.
 Regarding the aforementioned challenges and the potential of surrogate models, a benchmarking \textit{aerodynamic} dataset called \textit{AirfRANS} \cite{bonnet2022airfrans} was introduced to study the capabilities of DL in approximating the functional space of NS-PDEs (in this case, \textit{RANS} supplemented with the well-known k-$\omega$ SST turbulence model). This dataset is a model for real-world applications in aerodynamics, and the numerical solutions can be validated against experimental data available on the \textit{NASA Turbulence Modeling Resource} (TMR) of the \textit{Langley Research Center} \cite{TMR}. The simulations are run with \textit{OpenFoam} \cite{Jasak07openfoam:a,OpenFOAM}.
From a design standpoint, the challenges are: (1) simulations come in the form of unstructured mesh with millions of nodes, (2) High Reynolds leading to sharp signals, (3) Difficulty at encoding the geometry and boundary conditions w.r.t complex topological and physical variations including Angle of Attacks (AOA).
In this challenge, we propose to study the AirFoil design problem by considering a scarce data regime \cite{bonnet2022airfrans}. The task consists in predicting the incompressible steady-state two-dimensional fields and the force acting over airfoils in a subsonic regime. The goal is to find the airfoil that maximizes the lift-over-drag ratio and predict the velocity and pressure fields around it accurately. 
To \textit{physically} evaluate the DL models, only surface and volume fields are regressed.
Then, 
force coefficients are computed 
as post-treatments to stick with typical postprocessing of the \textit{RANS} equations. Therefore, the trained DL model is said to be \textit{physically consistent} only  if the predicted fields and the derived quantities predictions are consistent. Figure \ref{fig:cfd_gdl} illustrates an example of the input/outputs of CFD solvers and DL models.

\subsection{Metrics}


We propose an \textit{homogeneous evaluation} of   submitted solutions to learn  the airfoil design task  using the LIPS  platform. The evaluation is performed using 3 categories mentioned in Section \ref{sec: LIPS}: ML-Related, Physics compliance \& OOD generalization. 
For each category, specific criteria related to the airfoils design task are defined. 
The global score is computed based on linear combination of the scores related to three evaluation criteria categories:
\begin{equation}
\label{eq: score}
Score=\alpha_{ML}\times \bm{Score_{\textbf{ML}}} 
+ \alpha_{OOD}\times \bm{Score_{\textbf{OOD}}} + \alpha_{PH}\times \bm{Score_{\textbf{Physics}}},
\end{equation}
where $\alpha_{ML}$, $\alpha_{OOD}$ and $\alpha_{PH}$ are the coefficients to calibrate the relative importance of ML-Related, Application-based OOD, and Physics Compliance categories respectively.

We explain in the following subsections how each of the three sub-scores were calculated in the preliminary edition. As mentioned in section \ref{sec:novelty}, the evaluation criteria will slightly evolve during the new edition for a better  consideration of industrial applicability. 
\paragraph{ML-related Category score calculation}
This sub-score is calculated based on a linear combination of 2 sub-criteria: accuracy and speedup.
\begin{equation}
\label{eq: scoreML}
Score_{ML}=\alpha_A\times \bm{Score_{\textbf{Accuracy}}}
+ \alpha_S\times \bm{Score_{\textbf{Speedup}}},
\end{equation}
where $\alpha_A$ and $\alpha_S$ are the coefficients to calibrate the relative importance of accuracy and speedup respectively. For each quantity of interest, the accuracy sub-score is calculated based on two thresholds that are calibrated to indicate if the metric evaluated on the given quantity provides unacceptable/acceptable/great result. It corresponds to a score of  0 point / 1 point / 2 points, respectively. Within the sub-category, let: \added[id=ML]{$\sbullet[.75]$  \textcolor{red}{$Nr$}, the number of unacceptable results overall; $\sbullet[.75]$ \textcolor{orange}{$No$}, the number of acceptable results overall; $\sbullet[.75]$ \textcolor{ao(english)}{$Ng$}, the number of great results overall.} 
Let also $N$, given by $N=$ \textcolor{red}{$Nr$} $+$ \textcolor{orange}{$No$} $+$\textcolor{ao(english)}{$Ng$}. The score expression is given by
\begin{equation}
\label{eq: score_ML_accuracy}
Score_{\textbf{Accuracy}}= \frac{1}{2N} (2\times Ng+1\times No +0\times Nr) 
\end{equation}

A perfect score is obtained if all the given quantities provides great results. Indeed, we would have $N=Ng$ and $Nr=No=0$ which implies $Score_{\textbf{Accuracy}}=1$.

For the speed-up criteria, we calibrate the score using the $log_{10}$ function by using an adequate threshold of maximum speed-up to be reached for the task, meaning
\begin{equation}
\label{eq: score_ML_speed}
Score_{\textbf{Speedup}}=min\Biggl(\Biggl(\frac{log_{10}(SpeedUp)}{log_{10}(SpeedUpMax)}\Biggl),1\Biggl),
\end{equation}
where
\begin{itemize}
\item $SpeedUp$ is given by
\begin{equation}
\label{eq: speedUp}
SpeedUp=\frac{time_{\textbf{PhysicalSolver}}}{time_{\textbf{Inference}}},
\end{equation}
\item $SpeedUpMax$ is the maximal speedup allowed for the airfoil use case
\item $time_{\textbf{ClassicalSolver}}$: the elapsed time to solve the physical problem using the classical solver
\item $time_{\textbf{Inference}}$: the inference time.
\end{itemize}

In particular, there is no advantage in providing a solution whose speed exceeds $SpeedUpMax$, as one would get the same perfect score $Score_{\textbf{Speedup}}=1$ for a solution such that $SpeedUp=SpeedUpMax$.

Note that, while only the inference time appears explicitly in the score computation,   the training time is considered via  a fixed threshold:  if the training time overcomes 72 hours on a single GPU, the proposed solution will be rejected. Thus, its  global score is equal to zero.

\paragraph{Physical compliance category score calculation}
While the machine learning metrics are relatively standard, the physical metrics are closely related to the underlying use case and physical problem. 
There are two physical quantities considered in this challenge namely: the drag and lift coefficients.
For each of them, we compute two coefficients between the observations and predictions: \added[id=ML]{$\sbullet[.75]$ The spearman correlation, a nonparametric measure of the monotonicity of the relationship between two datasets (Spearman-correlation-drag : $\rho_{D}$ and Spearman-correlation-lift : $\rho_{L}$); $\sbullet[.75]$ The mean relative error (Mean-relative-drag : $C_D$ and Mean-relative-lift : $C_L$).}


For the Physics compliance sub-score, we evaluate the  relative errors of physical  variables. For each criteria, the score is also calibrated based on 2 thresholds and gives 0 /1 / 2 points, similarly to $score_{\textbf{Accuracy}}$, depending on the result provided by the metric considered. 

\paragraph{OOD generalization score calculation}
This sub-score will evaluate the capability on the learned model to predict OOD dataset. In the OOD testset, the input data are from a different distribution than those used for training. the computation of this sub-score is similar to $score_{ML}$ and is also based on two sub-criteria: accuracy and speed-up. To compute accuracy we consider the criteria used to compute the accuracy in $score_{ML}$ in addition to those  considered in physical compliance.

\paragraph{Score Calculation Example}

To demonstrate the calculation of the everall score, we utilize the notation established in the preceding section. We provide in table \ref{tab:Benchmark_evaluation_circle} several examples for the score calculation, including the top 5 solutions of the preliminary competition. We illustrate in the following how to calculate the score of the baseline (Fully connected) based on the parameters used in the preliminary edition: $\alpha_{ML}=0.4$,  $\alpha_{OOD}=0.3$, $\alpha_{PH}=0.3$, $\alpha_{A}=0.75$, $\alpha_{S}=0.25$,  $SpeedUpMax = 10000$.


\begin{itemize}
\item $Score_{\textbf{ML}} =  0.75\times(\frac{2\times 1 + 1 \times 1 + 0 \times 3 }{2\times 5})+ 0.25\times\frac{log_{10}(750)}{log_{10}(10000)}\approx0.405$
\item $Score_{\textbf{OOD}}=  0.75\times(\frac{2\times 1 + 1 \times 1 + 0 \times 7}{2\times 9})+ 0.25\times \frac{log_{10}(750)}{log_{10}(10000)}\approx0.305$
\item $ Score_{\textbf{Physics}}= (\frac{2\times 1}{2\times 4})=0.25$
\end{itemize}

For accuracy scores, the detailed results with their corresponding points are reported in Appendix \ref{append-tablescore}. Speed-up scores are calculated using the equation \ref{eq: speedUp} as follows:
\begin{itemize}
\item $time_{\textbf{PhysicalSolver}}= 1500s$, $time_{\textbf{Inference-ML}} = 2s$ , $time_{\textbf{Inference-OOD}} = 2s$ 
\item $ Score_{\textbf{SpeedupML}}  = Score_{\textbf{SpeedupOOD}} =  \frac{1500}{2} = 750$
\end{itemize}
Then, by combining them, the global score is $Score_{FC}=0.4\times 0.405+0.3\times 0.305+0.3\times 0.25=0.3285$, therefore 32.85\%.

\begin{table}[h]
    \centering
    \caption{Scoring table for Airfoil design task under 3 categories of evaluation criteria. The performances are reported using three colors computed on the basis of two thresholds. Colors meaning: \protect\myfullcircle{red} Unacceptable (0 point) \protect\myfullcircle{orange} Acceptable (1 point)  \protect\myfullcircle{green}  Great (2 points). Reported results: Baseline solution (Fully Connected NN), OpenFOAM (Ground Truth), and the 5 winning solutions from the preliminary edition of the Competition. MM-GP : Mesh morphing Gaussian Process. GGN-FC: a combined Graph Neural network (GNN) and FC appraoch. MINR: Multiscale Implici Neural Representations. Bi-Trans : Subsampled bi-transformer. NeurEco : NeurEco based MLP. }
    \resizebox{\columnwidth}{!}{
    \begin{tabular}{cccccccccccc}
    \toprule
         & & \multicolumn{9}{c}{\textbf{Criteria category}}\\
         & & \multicolumn{2}{c}{\textbf{ML-related (40\%)}} && \multicolumn{1}{c}{\textbf{Physics (30\%)}} && \multicolumn{3}{c}{\textbf{OOD generalization (30\%)}}  && \textbf{Global Score (\%)}\\ \cline{3-4} \cline{6-6} \cline{8-10} 
         & \textbf{Method} & Accuracy(75\%) & Speed-up(25\%) && Physical Criteria &&  OOD Accuracy(42\%) & OOD Physics(33\%) & Speed-up(25\%) &&\\ \cline{1-12}
         \multicolumn{1}{c|}{\multirow{8}{*}{}} & \multicolumn{1}{c|}{} & \hspace{-0.2cm}\underline{$\overline{u}_{x}$}\, \underline{$\overline{u}_{y}$}\, \underline{$\overline{p}$}\,\, \underline{$\overline{\nu}_{t}$}\,\,\, \underline{$\overline{p}_{s}$} & && \underline{$C_D$} \underline{$C_L$} \underline{$\rho_{D}$} \underline{$\rho_{L}$} &&  \hspace{-0.2cm}\underline{$\overline{u}_{x}$}\, \underline{$\overline{u}_{y}$}\, \underline{$\overline{p}$}\,\, \underline{$\overline{\nu}_{t}$}\,\,\, \underline{$\overline{p}_{s}$} & \underline{$C_D$} \underline{$C_L$} \underline{$\rho_{D}$} \underline{$\rho_{L}$}  & &&\\
         
         \multicolumn{1}{c|}{} & \multicolumn{1}{c|}{OpenFOAM} & \myfullcircle{green}\myfullcircle{green}\myfullcircle{green}\myfullcircle{green}\myfullcircle{green} & 1 &&   \myfullcircle{green}\myfullcircle{green}\myfullcircle{green}\myfullcircle{green}  &&  \myfullcircle{green}\myfullcircle{green}\myfullcircle{green}\myfullcircle{green}\myfullcircle{green} & \myfullcircle{green}\myfullcircle{green}\myfullcircle{green}\myfullcircle{green} &  1 && \textbf{82.5}\\
                  
         \multicolumn{1}{c|}{} & \multicolumn{1}{c|}{Baseline(FC)} & \myfullcircle{red}\myfullcircle{orange}\myfullcircle{red}\myfullcircle{green}\myfullcircle{red}  & 750 &&   \myfullcircle{red}\myfullcircle{orange}\myfullcircle{red}\myfullcircle{orange}  &&  \myfullcircle{red}\myfullcircle{orange}\myfullcircle{red}\myfullcircle{green} \myfullcircle{red} &\myfullcircle{red}\myfullcircle{red}\myfullcircle{red}\myfullcircle{red} &  750 && \textbf{32.85}\\ 

          \hline
          \hline
         Rank & \multicolumn{11}{|c}{Preliminary Edition : Top 5 solutions} \\
         
          \hline
          \hline
        \multicolumn{1}{c|}{1} & \multicolumn{1}{c|}{MMGP \cite{casenave2024mmgp}} &
        \myfullcircle{green}\myfullcircle{green}\myfullcircle{green}\myfullcircle{green}\myfullcircle{green} & 27.40 && \myfullcircle{green}\myfullcircle{green}\myfullcircle{orange}\myfullcircle{green}  &&  \myfullcircle{green}\myfullcircle{green}\myfullcircle{orange}\myfullcircle{green}\myfullcircle{orange} & \myfullcircle{green}\myfullcircle{green}\myfullcircle{orange}\myfullcircle{green}&  28.08 && \textbf{81.29}\\
          \cline{1-12}
                  \multicolumn{1}{c|}{2} & \multicolumn{1}{c|}{GNN-FC} &
        \myfullcircle{green}\myfullcircle{green}\myfullcircle{red}\myfullcircle{green}\myfullcircle{orange} & 570.77 &&   \myfullcircle{green}\myfullcircle{orange}\myfullcircle{orange}\myfullcircle{green}  &&  \myfullcircle{orange}\myfullcircle{green}\myfullcircle{red}\myfullcircle{orange}\myfullcircle{red} & \myfullcircle{green}\myfullcircle{orange}\myfullcircle{orange}\myfullcircle{orange}&  572.3 && \textbf{66.81}\\
          \cline{1-12}
                 \multicolumn{1}{c|}{3} & \multicolumn{1}{c|}{MINR} &
        \myfullcircle{green}\myfullcircle{green}\myfullcircle{orange}\myfullcircle{green}\myfullcircle{orange} & 518.58 && \myfullcircle{orange}\myfullcircle{orange}\myfullcircle{red}\myfullcircle{orange}  &&  \myfullcircle{green}\myfullcircle{green}\myfullcircle{red}\myfullcircle{green}\myfullcircle{red} & \myfullcircle{orange}\myfullcircle{orange}\myfullcircle{red}\myfullcircle{orange} &  519.21 && \textbf{58.37}\\
          \cline{1-12}
                \multicolumn{1}{c|}{4} & \multicolumn{1}{c|}{Bi-Trans } &
        \myfullcircle{green}\myfullcircle{green}\myfullcircle{red}\myfullcircle{green}\myfullcircle{red} & 552.97 && \myfullcircle{orange}\myfullcircle{orange}\myfullcircle{red}\myfullcircle{green}  &&  \myfullcircle{orange}\myfullcircle{orange}\myfullcircle{red}\myfullcircle{orange}\myfullcircle{red} & \myfullcircle{orange}\myfullcircle{red}\myfullcircle{red}\myfullcircle{orange}&  556.46 && \textbf{51.24}\\
          \cline{1-12}
                        \multicolumn{1}{c|}{5} & \multicolumn{1}{c|}{NeurEco  } &
        \myfullcircle{green}\myfullcircle{green}\myfullcircle{orange}\myfullcircle{green}\myfullcircle{red} & 44.93 && \myfullcircle{orange}\myfullcircle{orange}\myfullcircle{red}\myfullcircle{orange}  &&  \myfullcircle{green}\myfullcircle{green}\myfullcircle{red}\myfullcircle{green}\myfullcircle{red} & \myfullcircle{orange}\myfullcircle{orange}\myfullcircle{red}\myfullcircle{orange} &  44.78 && \textbf{50.72} \\
         \bottomrule
    \end{tabular}}
    \label{tab:Benchmark_evaluation_circle}
    \vspace{-.5cm}

\end{table}

\subsection{Baselines, code, and material provided}


The provided starting kit\footnote{\href{https://github.com/IRT-SystemX/ml4physim\_startingkit}{https://github.com/IRT-SystemX/ml4physim\_startingkit}} comprises a series of Jupyter notebooks designed to assist participants in getting started with the airfoil simulation and how to contribute to the competition. It also features a fully documented implementation of the baseline solution, aiding participants in replicating baseline results. This starting kit was utilized in the preliminary edition of the competition and has been continuously enhanced. The list of available notebooks is outlined in Appendix A.

Furthermore, the code for the top 5 solutions from the preliminary edition will be accessible in a dedicated repository prior to the commencement of this competition.


\subsection{Website, tutorial and documentation}

The competition website will be based on the one used for the preliminary iteration  \footnote{\href{https://ml-for-physical-simulation-challenge.irt-systemx.fr/airfoil-challenge-1}{https://ml-for-physical-simulation-challenge.irt-systemx.fr/airfoil-challenge-1}}. It aims to serve as a central hub for all essential information, including:
\begin{enumerate*}[label=(\roman*)]
\item General details
\item Organization, rules, and regulations
\item Links and instructions for the submission platform (Codabench)
\item Announcements and recordings of webinars
\item Tutorial section with links to relevant parts of the starting kit repository
\item Contact information for the organizers
\end{enumerate*}

Interactive webinars will also be conducted and recorded to offer insights into the competition, airfoil design simulation, and submission process on Codabench. Additionally, we have set up a dedicated email address and Discord channel to facilitate open communication with all participants (or potential participants). During the preliminary edition of the challenge, assistance and information were provided via Discord, and we remained available to help all participants.





\section{Organizational aspects}
\subsection{Protocol}
The competition will be run on the Codabench platform \cite{xu2021codabench}. The link will be provided on the competition website. Participants will have to: 1) create an account; 2) download a starting kit to prepare their submission; 3) upload on the Codabench platform some code compliant with the described interfaces. Then, the submissions will be trained on the dedicated GPU cluster and the LIPS framework will be used  to evaluate the submissions and compute the global score. The score will be published on the Codabench competition page and the participants will also have access to an additional page with the detailed metrics. In order to prevent overfitting, the submissions will be evaluated also on other test sets that are different from ones provided in the dataset.

\subsection{Rules and Engagement}

This challenge starts on 10 June  2024 and ends on 17th October 2024.
\begin{itemize}[nolistsep,leftmargin=*]

    \item This challenge is open to anyone and runs in 3 phases;
    \item During Phase 1 and Phase 2, the participants can submit their codes and see their results on the leader board;
   \item During the warm-up phase, the organizers may adjust the global score formula;
  \item During phase 1 and 2 the participants can submit their codes and see their scores on the leader board;
\item The organizers may provide additional baseline results during the challenge to stimulate the competition;
\item The participant, and/or the team using a group account, will be limited to 10 submissions per day and 500 in total per phase;
\item The organizers strongly encourage all participants to share their codes and make them accessible in public submission;
\item The final ranking of the participants will be performed using the global score calculated based on the 3 categories of criteria, and notified to all participants;

    \item Teams should use a common account, under a group email. Multiple accounts are forbidden;
       \item To receive any price, a team should agree to open-source its code at the end of the competition. Deadline : 2 weeks before the event at Neurips. 

\end{itemize}



\textbf{Communication:} 
The organizers will announce each phase (beginning or end) using the challenge mailing list; a forum (ex. using a Discord channel) will be provided to ensure that participants can contact the organizers to ask any questions at any time regarding the rules and the overall organization. Bug reports can be reported as issues on the LIPS \& AirfRANS Github pages.
\subsection{Schedule and readiness}

\label{sec:schedule}
\paragraph{Detailed timeline}

The competition will consist of three phases.

Firstly, the \textit{warm-up phase} allows participants to familiarize themselves with all provided materials and the competition platform. They can make initial submissions and offer feedback to organizers. Organizers will use this feedback to adjust and refine the competition setup for the subsequent phase.

In the \textit{Development phase}, contestants will have the opportunity to test and enhance their models using the provided validation datasets. Throughout this phase, participants can train and refine their models using their own resources, leveraging materials from the starting kit. To validate a score, participants must submit their model on the Codabench platform, where it will be retrained using the competition resources.

Finally, in the \textit{Final phase}, organizers will validate the rankings obtained at the end of the development phase. This validation process involves verifying the code of submitted models and assessing the robustness of the best solutions through multiple trials. All submitted models will undergo retraining on our servers and evaluation using a private test dataset similar to the one provided during the development phase.

The proposed schedule is the following:


\begin{itemize}
\item {\bf  Competition period (July 1st - October 31st)}   {\bf Warm up phase  :  5 weeks}, {\bf Development phase :  10 weeks} , {\bf Final phase:  4 weeks}. 




\item Announcement of results: November 10th, 2024.
\item Fact sheets and code release by winners due: November 25th, 2024.
\item Presentation of results at NeurIPS competition workshop: December 9th 2024.
\end{itemize}

\vspace{-0.5cm}
\subsection{Competition promotion and incentives}
The challenge will be advertised through different channels:
announcements in major conferences (NeurIPS, ICLR, ICML) and related workshops about ML and Physical Science;
mailing lists, social networks;
NVIDIA blog. Monetary awards will be offered to the winners of each task of the competition: 1st place 4000 EUR, 2nd place 2000 EUR, 3rd place 1000 EUR. Finally, a webinar presentation of winning solutions will be organized, in which participants will have the opportunity to present in depth their solution.
They will also be invited to write a joint paper about the competition results.


\vspace{-0.3cm}
\section{Resources}
\vspace{-0.2cm}
\subsection{Resources provided by organizers}


In order to evaluate contestant submissions, a dozen of GPUs A6000 will be sponsored by NVIDIA and made them available through Exaion Infrastructure in Paris-Saclay France to power compute workers connected to Codabench. This connection is already available and has served to train \& evaluate  the  submissions received  for the preliminary edition of the competition (between November 2023 \& March 2024). 

\subsection{Support requested}

We would be very grateful if we could beneficiate from the support of the NeurIPS 2023 Competition Track organizers in particular in promoting our challenge via their own channels.

\bibliographystyle{unsrt}
\bibliography{references}

\subsection{Organizing team}
The team organizing this challenge is composed of researchers and engineers in the domains of : Deep and Machine learning, Scientific Computing, Software Programming and High-Performance Computing, as well as Numerical Simulation and Computational Fluid Dynamics. The short biography of each member is provided bellow. 

The organizers extend their gratitude to \textbf{Isabelle Guyon }for her valuable contributions to the discussions and insights regarding the proposed challenge.

\parpic{\includegraphics[width=0.6in,clip, keepaspectratio]{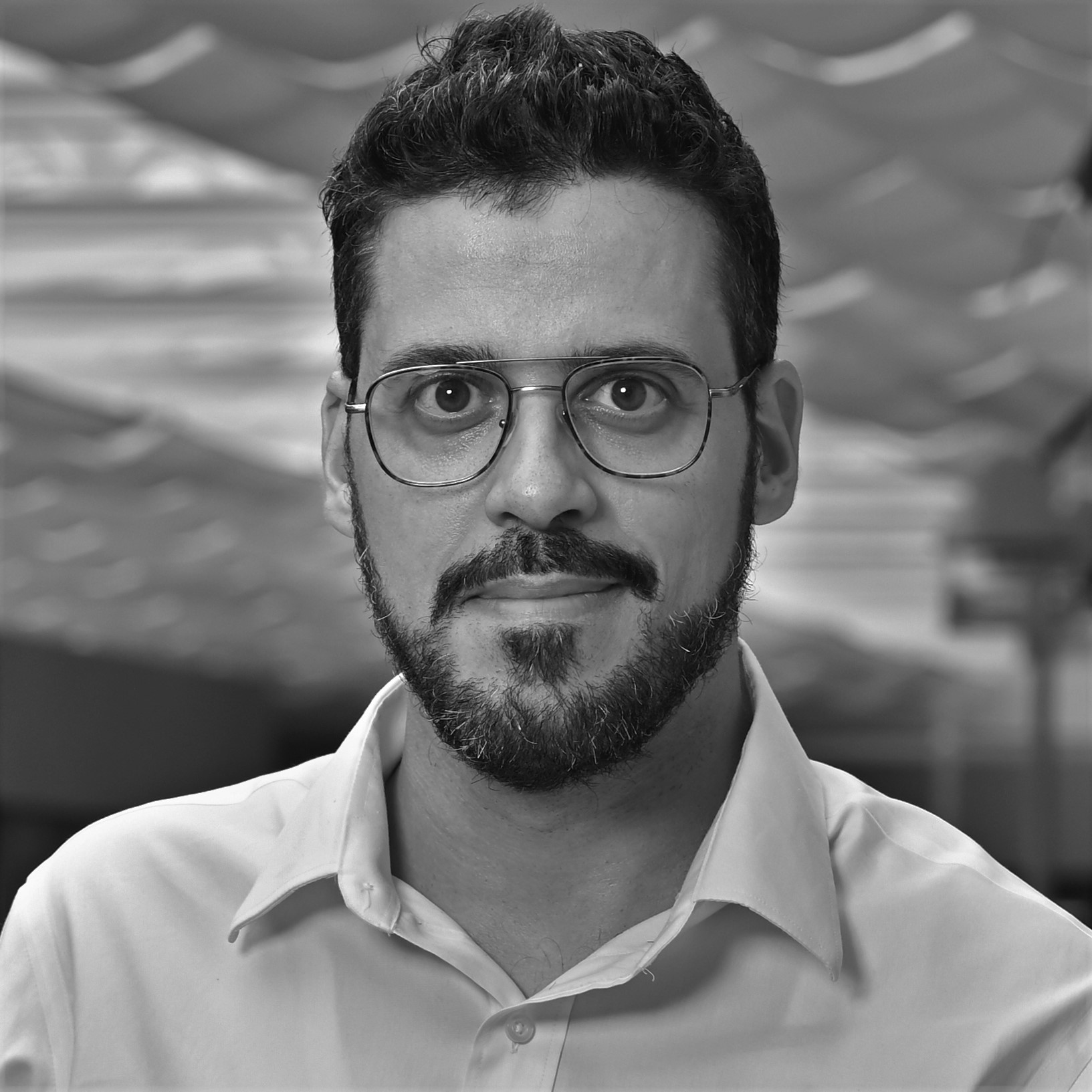}}
\noindent {\bf Mouadh Yagoubi (Leader organizer, Baseline method provider, Evaluator)} is a Senior Researcher at IRT SystemX working on hybridization of physical simulation and AI. He  received a Ph.D in applied mathematics from INRIA Saclay  in 2012. His research interest includes  machine learning, evolutionary computation and ML based surrogate modeling  to solve computationally expensive physical problems. 

\vspace{0.6cm}
\parpic{\includegraphics[width=0.6in,clip,keepaspectratio]{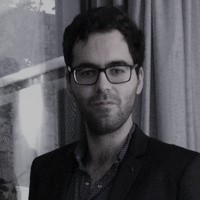}}
\noindent {\bf David Danan (Baseline method provider, Evaluator)} is a research engineer at IRT SystemX in the scientific calculation and optimization team. He received a Ph.D. in applied mathematics to contact mechanics from UPVD in France in 2016. His current research topics includes topology optimization, solid mechanics and hybridization between PDE solver-based approaches and machine learning techniques.

\parpic{\includegraphics[width=0.6in,clip,keepaspectratio]{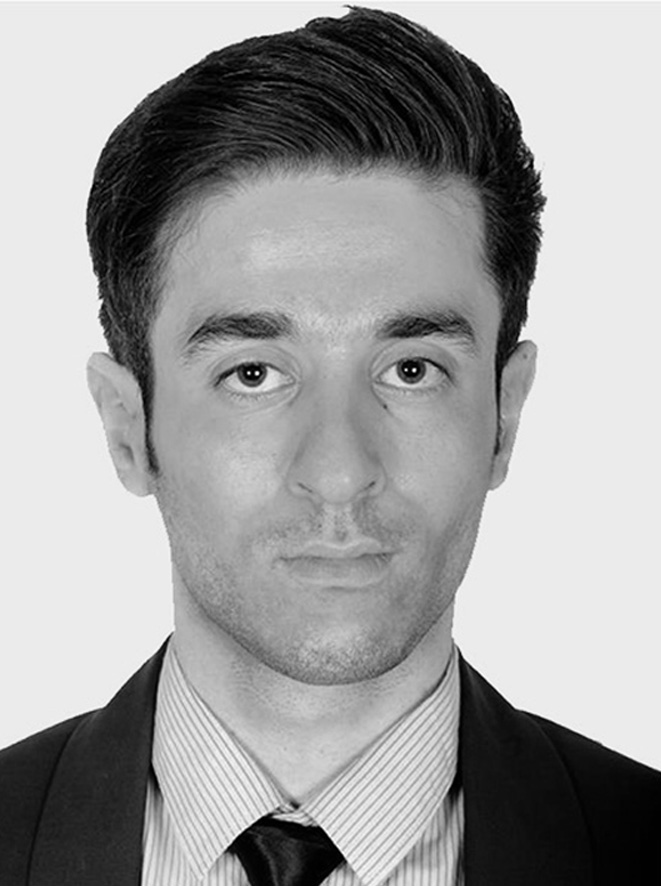}}
\noindent {\bf Milad Leyli-abadi (LIPS platform administrator, Evaluator)} received a Ph.D. in machine learning and applied mathematics from Paris-Est Cr\'{e}teil University (UPEC) in 2019. He is currently research engineer at IRT SystemX in France and is working on problems related to hybridization of physical simulation and AI. His research interests include time series modeling, forecasting models, machine learning and big data analytics. 
\vspace{0.6cm}

\parpic{\includegraphics[width=0.6in,clip,keepaspectratio]{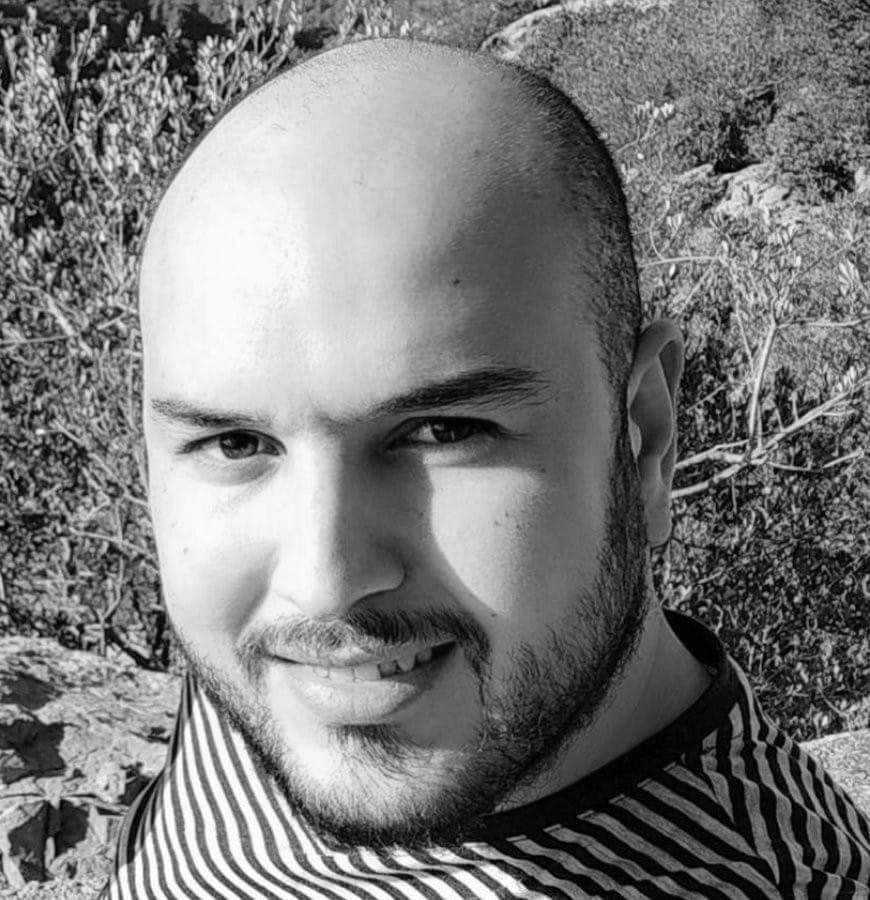}}
\noindent {\bf Jocelyn Ahmed Mazari (Data provider, Baseline method provider, Evaluator)} is a senior deep learning researcher at SimAI team, Ansys Inc, working on deep learning models for computational fluid dynamics. His research interests include geometric deep learning, multi-scale modeling, partial differential equations, and statistical signal processing. He received a Ph.D. in deep learning and computer vision from Sorbonne Université (Campus Pierre et Marie Curie) in 2020.

\vspace{0.6cm}
\parpic{\includegraphics[width=0.6in,clip,keepaspectratio]{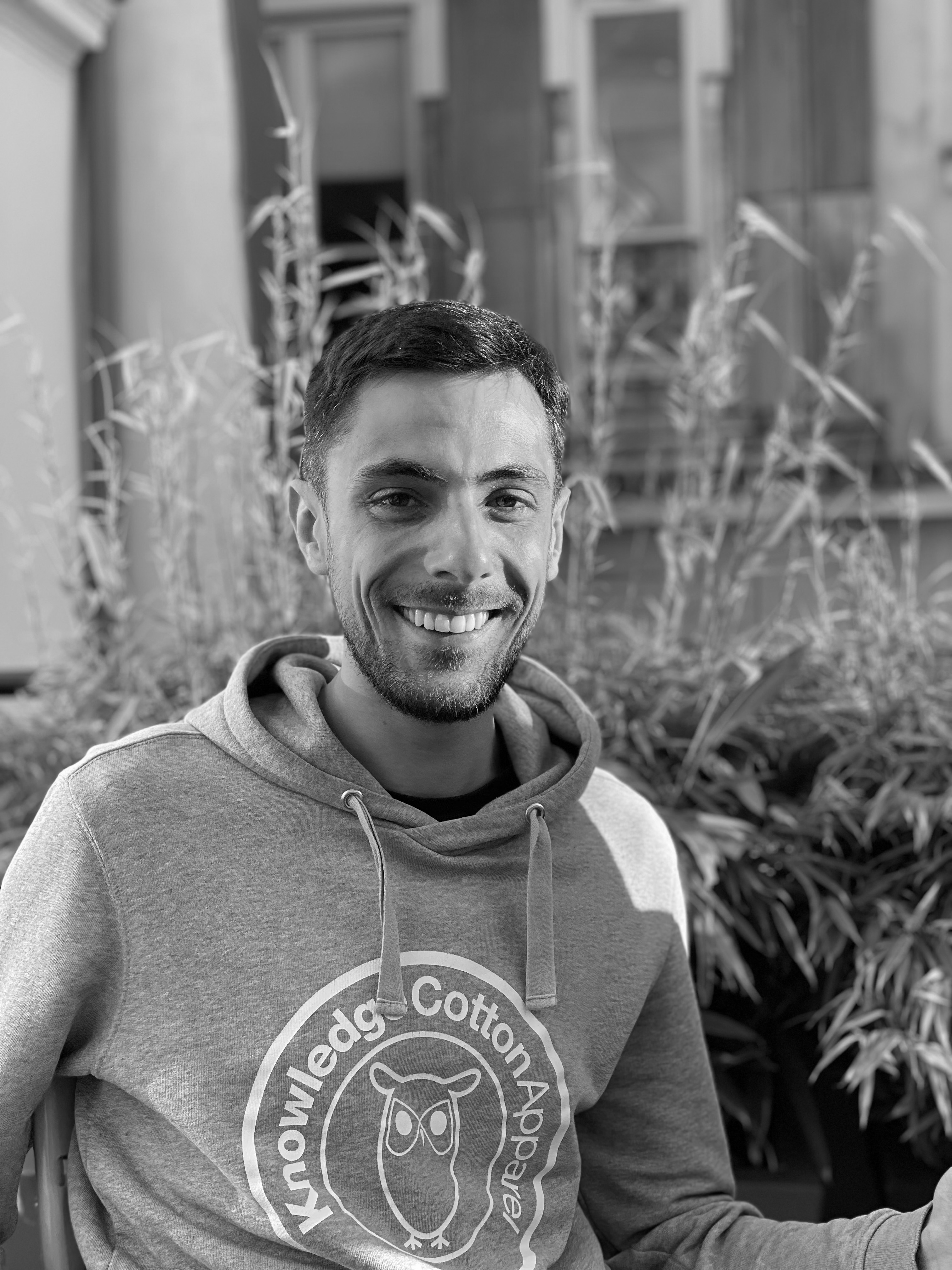}}
\noindent {\bf Florent Bonnet (Data provider, Baseline method provider)} is a second-year Ph.D. student at Sorbonne Université (MLIA team within ISIR laboratory) and SimAI team, Ansys Inc. He is co-advised by Professor Patrick Gallinari (thesis director and head of the MLIA team) and Jocelyn Ahmed Mazari (industrial advisor at SimAI team, Ansys Inc). His research focuses on understanding and building physically constrained deep learning models to solve partial differential equations. He holds a Master degree from École normale supérieure Paris-Saclay (ENS Paris-Saclay) in machine learning (Master MVA).

\vspace{0.6cm}
\parpic{\includegraphics[width=0.6in,clip,keepaspectratio]{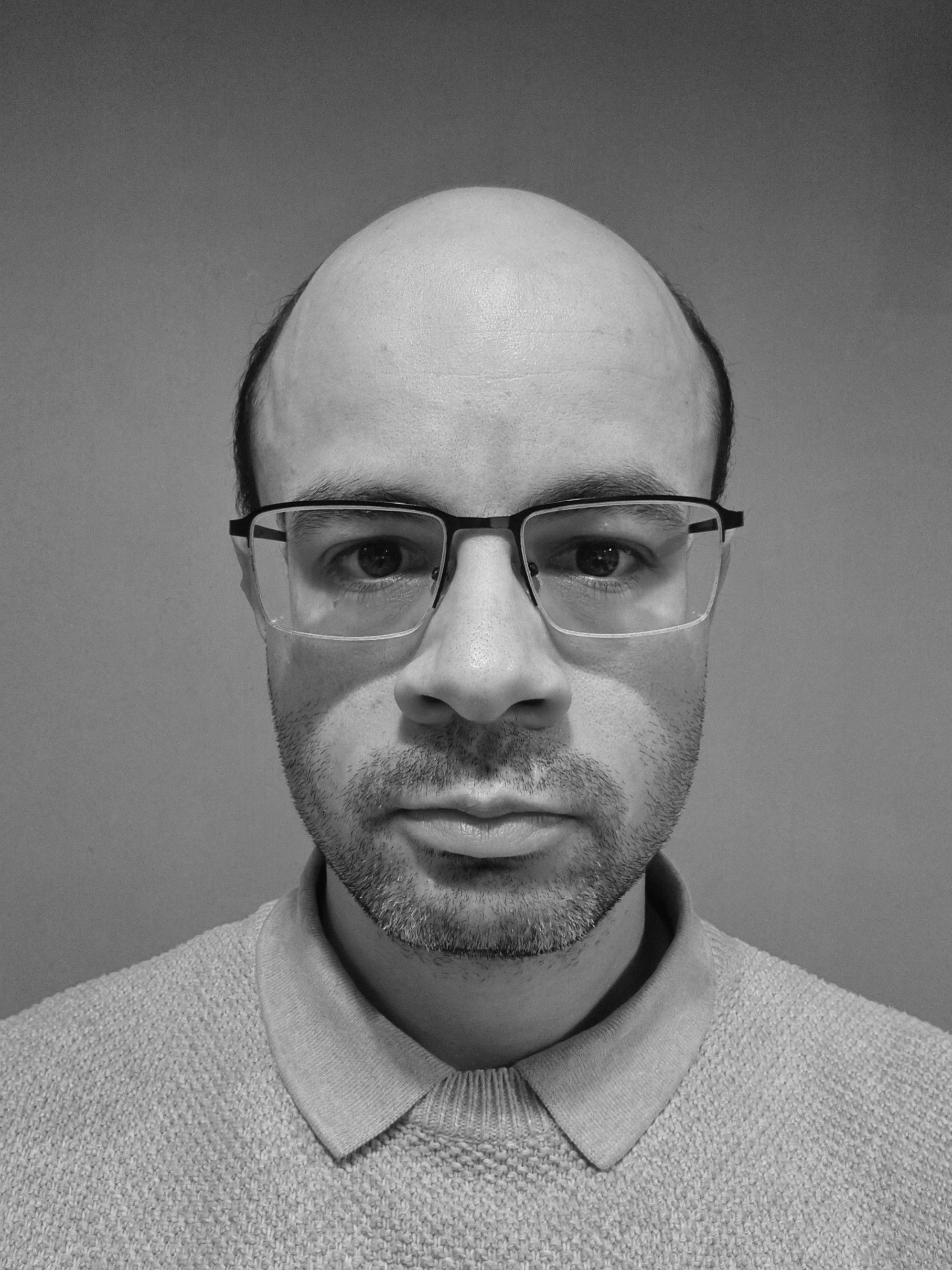}}
\noindent {\bf Jean-Patrick Brunet (Codabnch Platform administrator, Evaluator)} Jean-Patrick is a software architect at the Technological Research Institute SystemX. Holding two masters in mechanical and petroleum engineering from the Ecole Centrale Nantes and Penn State University (2013). His research interest are high performance computing and collaborative engineering solutions.

\vspace{0.6cm}
\parpic{\includegraphics[width=0.6in,clip,keepaspectratio]{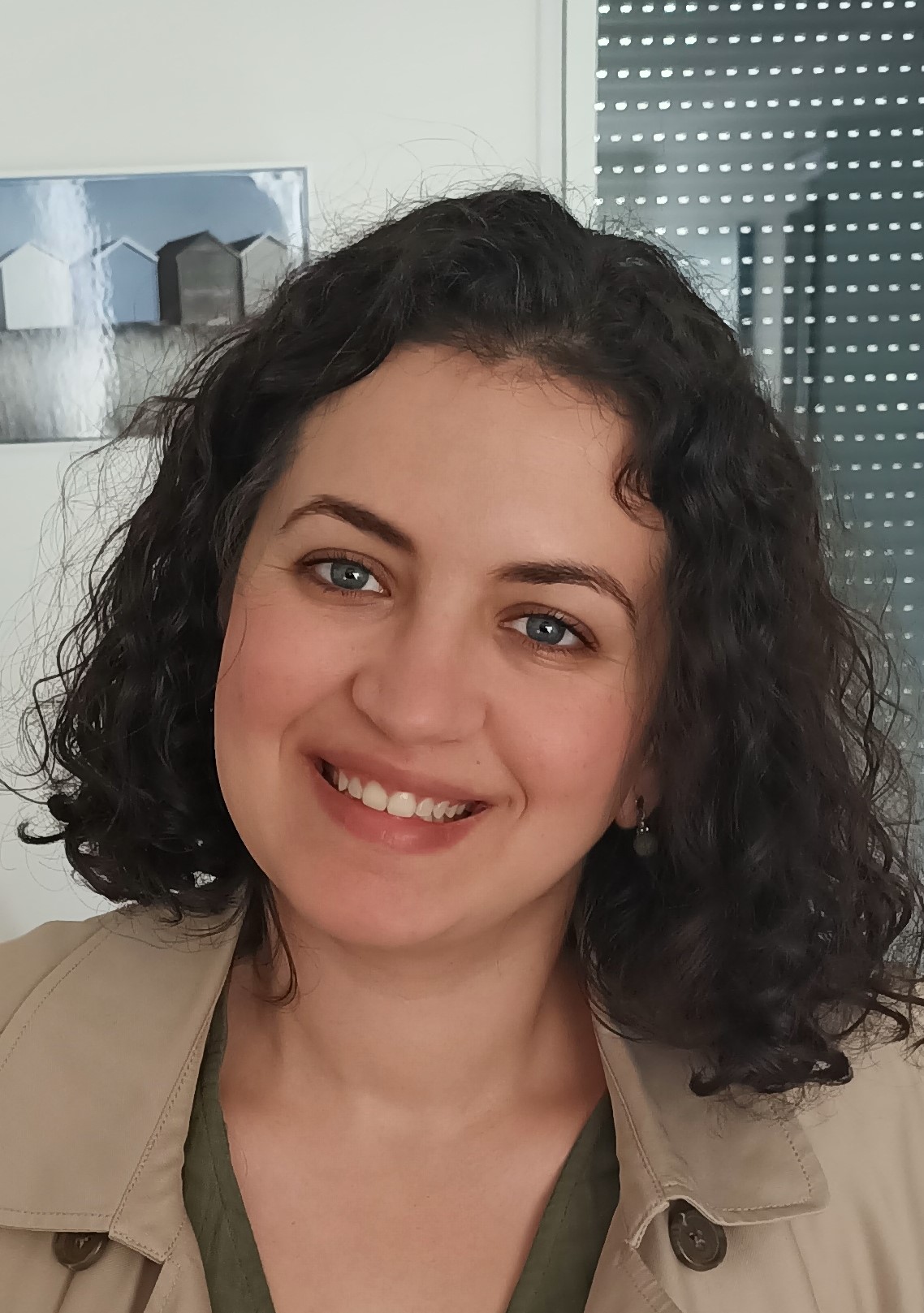}}
\noindent {\bf Maroua Gmati (Competition website and LIPS  developer)} is a software engineer at IRT SystemX, specializes in crafting software solutions. Her expertise spans both front-end and back-end development, allowing her to play a pivotal role in creating comprehensive solutions. She completed her Master of Science in Informatics from Université Grenoble Alpes, following her earlier achievement of an engineering degree in computer science.

\vspace{0.6cm}
\parpic{\includegraphics[width=0.6in,clip,keepaspectratio]{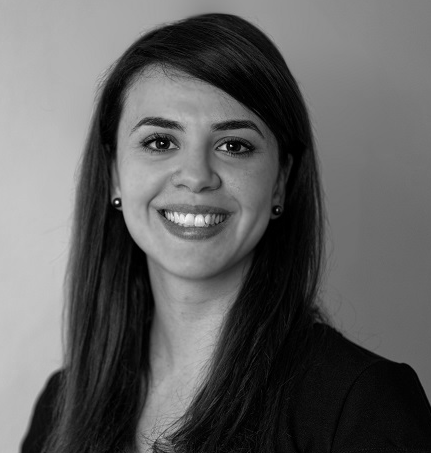}}
\noindent {\bf Asma Farjallah (Infrastructure provider, evaluator)} is senior solutions architect at NVIDIA, helping customers in the energy industry accelerate their HPC and deep learning workloads using GPUs. Prior to joining NVIDIA, Asma worked as an application engineer at Intel where she helped optimize scientific workloads on Intel's technologies. She holds a PhD in computational science from the University of Versailles Saint-Quentin en Yvelines. 

\vspace{0.6cm}
\parpic{\includegraphics[width=0.6in,clip,keepaspectratio]{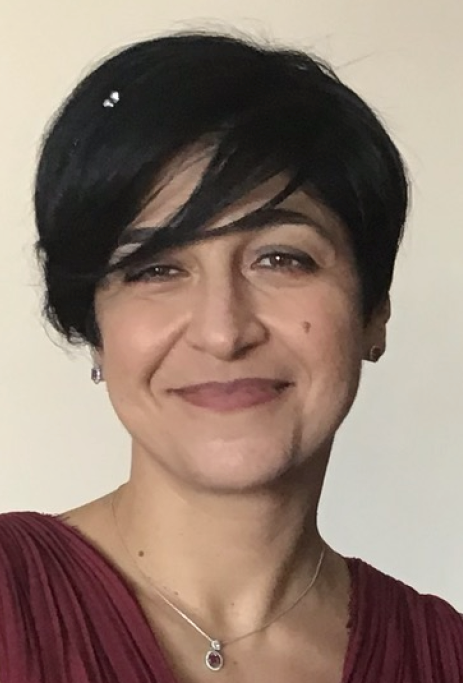}}
\noindent{\bf Paola Cinnella (Scientific Advisor) } is a Professor of Fluid Dynamics at Sorbonne University in Paris. Her research covers several facets of Computational Fluid Dynamics (CFD), including high-order methods, data-driven and machine-learning assisted approaches for turbulent flow modeling, shape optimization, uncertainty quantification, and  applications of CFD to the analysis and design of  flows. She is editor-in-chief of the international journal \emph{Computers \& Fluids}, associate editor of the \emph{International Journal of Heat and Fluid Flow}, and editorial board member of \emph{Flow, Turbulence and Combustion}.
She is also a member of the Aerodynamics Panel of the French Association of Aeronautics and Astronautics and the scientific coordinator of the scientific interest group in Machine Learning for Fluid Dynamics for the European Research Community on Flow, Turbulence and Combustion (ERCOFTAC). \\

\vspace{0.6cm}
\parpic{\includegraphics[width=0.6in,clip,keepaspectratio]{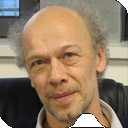}}
\noindent{\bf 
Patrick Gallinari (Scientific Advisor)} is a professor in Computer Science at Sorbonne University in Paris. His research focuses on statistical learning with applications in different fields such as semantic data analysis and complex data modeling. He discovered the ML domain in the mid 80es when he started to work on Neural Networks, an emerging field at that time. He has been one of the pionneers of this research domain in France/ Europe and worked on NN and on other ML models since that. He investigated different application domains like Information Retrieval, Social Data analysis, User Modeling. Today his main focus is on Physics Aware Deep Learning and on some aspects of Natural Language Processing. He has been leading the Machine Learning team MLIA for some years. He has been director of the computer science lab. at Sorbonne University (LIP6) for 9 years (2005 to 2013) and vice director for 6 years before, He also acted as vice director of the scientific committee of the faculty of engineering at UPMC (2010 to 2021).\ \

\vspace{0.6cm}
\parpic{\includegraphics[width=0.6in,clip,keepaspectratio]{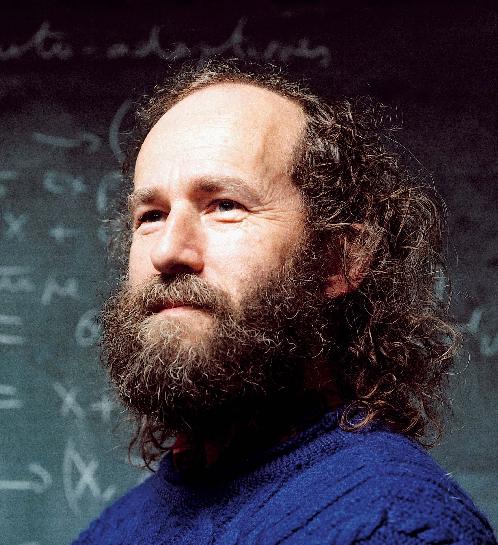}}
\noindent{\bf 
Marc Schoenauer (Scientific Advisor) } is Principal Senior Researcher (DR0) with INRIA, that he joined in 2001 after 20 years with CNRS, at CMAP of Ecole Polytechnique. He founded the TAO team at INRIA Saclay together with Michèle Sebag in 2003. He has been working at the border between Evolutionary Computation (EC) and Machine Learning (ML), author of more than 150 papers, (co-)advisor of 35 PhD students. He has been Chair of ACM-SIGEVO (2015-2019), was the founding president (1995-2002) of Evolution Artificielle, and president of AFIA (2002-2004). He has been Editor in Chief of Evolutionary Computation Journal (2002-2009, now on the Advisory Board), is or has been in the Editorial Board of the most prestigious journals in EC, and since 2013 is Action Editor of JMLR. He seconded Cédric Villani in writing his report on the French Strategy for AI delivered to Pdt Macron in March 2018, and is currently Deputy Research Director in charge of AI at INRIA.\ \


\appendix
\newpage
\section{Starting kit description}

\begin{itemize}
    \item \verb|0-Basic_Competition_Information|: This notebook contains general information concerning the competition organization, phases, deadlines and terms. The content is the same as the one shared in the competition Codabench page. 
    \item \verb|1-Airfoil_design_basic_simulation|: This notebook aims to familiarize the participants with the use case and to facilitate their comprehension. It allows the visualization of some simulation results. 
    \item \verb|2-Import_Airfoil_design_Dataset|: Shows how the challenge datasets could be downloaded and imported using proper functions. These data will be used in the following notebook to train and evaluate an augmented simulator. 
    \item \verb|3-Reproduce_baseline_results|: This notebook shows how the baseline results could be reproduced. It includes the whole pipeline of training, evaluation and score calculation of an augmented simulator using LIPS platform. 
    \item \verb|3b-Reproduce_baseline_results_Advanced_Configuration|: This notebook shows how another baseline results could be reproduced. It  also includes the whole pipeline of training, evaluation and score calculation of an augmented simulator using LIPS platform.
    \item \verb|4-How_to_Contribute|: This notebook shows 3 ways of contribution for beginner, intermediate and advanced users. The submissions should respect one of these forms to be valid and also to enable their proper evaluation through the LIPS platform which will be used for the final evaluation of the results. 
    
    \begin{itemize}
        \item Beginner Contributor: You only have to calibrate the parameters of existing augmented simulators
        \item Intermediate Contributor: You can implement an augmented simulator respecting a given template (provided by the LIPS platform)
        \item Advanced Contributor: you can implement your architecture independently from LIPS platform and use only the evaluation part of the framework to assess your model performance.
    \end{itemize}
    \item \verb|4a-How_to_Contribute_Tensorflow|: This notebook shows how to contribute using the existing augmented simulators based on Tensorflow library. The procedure to customize the architecture is fairly the same as pytorch (shown in Notebook 4).
    \item \verb|5-Scoring|: This notebook shows firstly how the score is computed by describing its different components. Next, it provides a script which can be used locally by the participants to obtain a score for their contributions. We encourage participants to evaluate their solutions via codabench (which uses the same scoring module as the one described in this notebook).
     \item \verb|6-Submission|: This notebook presents the composition of a submission bundle for Codabench and usable parameters.
     \item \verb|7-Submission_examples|: This notebook shows how to submit on Codabench and examples of submissions bundles.
    \\
    \\
\end{itemize}

\section{Detailed results of the baseline solution (fully connected Neural Network)}
\label{append-tablescore}

\begin{table}[H]
    \centering
    \vspace{-0.9cm}
    \caption{Accuracy scores calculation of the FC solution.}
    \begin{tabular}{cccccc}
    \toprule
    Category & Criteria & obtained results & Thresholds & min/max & obtained score \\

     \cline{1-6} 
     \addlinespace
     \multirow{5}{*}{ML-Related} & $\overline{u}_{x}$\ & 0.208965 & T1=0.1 / T2 =0.2 & min & \myfullcircle{red}  0 point \\
     & $\overline{u}_{y}$\ & 0.144508 &  T1=0.1 / T2=0.2 & min & \myfullcircle{orange}  1 point \\
     & $\overline{p}$\ & 0.193066 & T1=0.02 / T2=0.1 & min & \myfullcircle{red}  0 point \\ 
     & $\overline{\nu}_{t}$\ & 0.277285 & T1=0.5 / T2=1.0 & min & \myfullcircle{green}  2 points  \\
     & $\overline{p}_{s}$ & 0.425576 & T1=0.08 / T2 =0.2  & min& \myfullcircle{red}  0 point  \\
     \cline{1-6} 
          \addlinespace
      \multicolumn{6}{r}{$N$ = 5, $Nr$ = 3, $No$ = 1, $Ng$ = 1.   } \\
    \toprule 
    \toprule
        \multirow{4}{*}{Physical compliance } 
     
     & $C_D$ &16.345740 & T1=1 / T2 =10  & min& \myfullcircle{red}  0 point  \\
     & $C_L$ & 0.365903 & T1=0.2 / T2 =0.5  & min& \myfullcircle{orange}  1 point  \\
     & $\rho_{D}$ & -0.043079 & T1=0.5 / T2 =0.8  & max& \myfullcircle{red}  0 point  \\
     & $\rho_{L}$ & 0.957070 & T1=0.94 / T2 =0.98  & max& \myfullcircle{orange}  1 point  \\
     \cline{1-6} 
          \addlinespace
      \multicolumn{6}{r}{$N$ = 4, $Nr$ = 2, $No$ = 2, $Ng$ = 1.  } \\
       \toprule
        \toprule
         \addlinespace
     \multirow{9}{*}{OOD Generalization} & $\overline{u}_{x}$\ & 0.322766 & T1=0.1 / T2 =0.2 & min & \myfullcircle{red}  0 point \\
     & $\overline{u}_{y}$\ & 0.199635 &  T1=0.1 / T2=0.2 & min & \myfullcircle{orange}  1 point \\
     & $\overline{p}$\ & 0.333169 & T1=0.02 / T2=0.1 & min & \myfullcircle{red}  0 point \\ 
     & $\overline{\nu}_{t}$\ & 0.431288 & T1=0.5 / T2=1.0 & min & \myfullcircle{green}  2 points  \\
     & $\overline{p}_{s}$ & 0.805426 & T1=0.08 / T2 =0.2  & min& \myfullcircle{red}  0 point  \\
     & $C_D$ & 21.793367 & T1=1 / T2 =10  & min& \myfullcircle{red}  0 point  \\
     & $C_L$ & 0.711271 & T1=0.2 / T2 =0.5  & min& \myfullcircle{red}  0 point  \\
     & $\rho_{D}$ & -0.043979 & T1=0.5 / T2 =0.8  & max& \myfullcircle{red}  0 point  \\
     & $\rho_{L}$ & 0.917206 & T1=0.94 / T2 =0.98  & max& \myfullcircle{red}  0 point  \\
     \cline{1-6} 
          \addlinespace
      \multicolumn{6}{r}{$N$ = 9, $Nr$ = 7, $No$ = 1, $Ng$ = 1.  } \\
        \toprule
        \toprule
             \addlinespace
    \end{tabular}
    \label{tab:detailed_results}

\end{table}

\end{document}